\documentclass[]{elsart}
\usepackage{graphicx}
\usepackage{amsmath}
\usepackage{delarray}
\usepackage{natbib}

\def\ba{\mbox{\boldmath $a$}}
\def\br{\mbox{\boldmath $r$}}

\def\bx{\mbox{\boldmath $x$}}
\def\bv{\mbox{\boldmath $v$}}

\def\bA{\mbox{\boldmath $A$}}

\def\badot{\mbox{\boldmath $\dot a$}}
\def\batwo{\mbox{\boldmath $\ddot a$}}

\def\bj{\mbox{\boldmath $j$}}
\def\bs{\mbox{\boldmath $s$}}
\def\bc{\mbox{\boldmath $c$}}
\def\bA{\mbox{\boldmath $A$}}
\def\bJ{\mbox{\boldmath $J$}}
\def\bS{\mbox{\boldmath $S$}}
\def\bC{\mbox{\boldmath $C$}}

\newcommand{\dt}{\Delta t}

\begin{document}
\begin{frontmatter}
\title{6th and 8th Order Hermite Integrator for $N$-body Simulations}
\author[Hongo]{Keigo Nitadori} 
\author[NAOJ]{Junichiro Makino}
\address[Hongo]{Department of Astronomy, University of Tokyo, 7-3-1 Hongo, Bunkyo-ku, Tokyo 113-0033, Japan}
\address[NAOJ]{National Astronomical Observatory of Japan, Mitaka, Tokyo, 181-8588, Japan }

\begin{abstract}
We present sixth- and eighth-order Hermite integrators for
astrophysical $N$-body simulations, which use the derivatives of
accelerations up to second order ({\it snap}) and third order ({\it
crackle}).  These schemes do not require previous values for the
corrector, and require only one previous value to construct the
predictor.  Thus, they are fairly easy to implement. The additional
cost of the calculation of the higher order derivatives is not very
high. Even for the eighth-order scheme, the number of floating-point
operations for force calculation is only about two times larger than that
for traditional fourth-order Hermite scheme.
The sixth order scheme is better than the traditional fourth order
scheme for most cases. When the required accuracy is very high, the
eighth-order one is the best.
These high-order schemes have several practical advantages. For
example, they allow a larger number of particles to be integrated in
parallel than the fourth-order scheme does, resulting in higher
execution efficiency in both general-purpose parallel computers and
GRAPE systems.

\end{abstract}


\end{frontmatter}

\section{Introduction}

\cite{Aarseth63} introduced what is now called the ``Aarseth scheme'' or
``the standard scheme'' for the direct integration of gravitational
$N$-body systems. It is a combination of the individual timestep
algorithm, which allows individual particles to have their own times
and timesteps, and variable-stepsize fourth-order Adams-Moulton
predictor-corrector scheme.

The basic idea of individual timestep scheme is as follows. When
particle $i$ is integrated from its time $t_i$ to its new time $t_i +
\Delta t_i$, the calculation of acceleration is done only at its new
time, and positions of all other particles at that time are
``predicted'' in some way. Thus, Adams-Moulton predictor-corrector
schemes in the PEC (predict-evaluate-correct) mode with variable
stepsize are suitable for the individual timestep algorithm, because
they require the acceleration calculation only at the end of the
timestep. In addition, in the PEC mode, the acceleration can be
calculated from predicted variables.

If we do not use the individual timestep algorithm, we can easily
change timesteps if we use single-step integration schemes such as
Runge-Kutta methods.  However, Runge-Kutta
schemes cannot be combined with the individual timestep algorithm,
because they require the calculation of accelerations in intermediate
points. In the case of two particles with different timesteps, in order
to integrate the particle with longer timestep, we need the position
of the other particle in the past. However, with usual implementation
of the individual timestep algorithm, such past data is not
available. In principle, we could keep the past trajectory of
particles as demonstrated by \cite{MHKS06}. Such schemes are
not yet widely used, though a sample implementation does
exist\citep{HMACS}.

The fourth-order Aarseth scheme had been the method of choice for the time
integration of gravitational $N$-body systems. However, the optimal
value for the order of the integration scheme has not been known.
\cite{Makino91a} implemented the Aarseth scheme with an
arbitrary order, and performed a systematic test of the accuracy. He
found that the optimal choice of the order weekly depends on the
required accuracy, and if the required accuracy is very high orders
higher than 4 would give better results. However, he also found that
the fourth-order scheme is close to optimal for practical values of
required accuracy. His result, however, is for a pure individual
timestep algorithm, for which the calculation cost of the acceleration
depends on the order of the integrator, through the calculation cost
of predictors for particles other than that integrated.
\cite{McMillan86} and later \cite{Makino91b} introduced the so-called
blockstep scheme, in which the timesteps of particles are quantized to
powers of two so that multiple particles share exactly the same
time. With this blockstep scheme, the calculation cost of predictors
becomes much smaller than that of the force calculation for any
practical value of the order of the integration scheme, and therefore
high-order schemes become more efficient than in the case of the
original individual timestep algorithm.

\cite{Makino91a} also introduced the
concept of Hermite scheme, in which the first time derivative of the
acceleration is directly calculated and used to construct the
interpolation (actually the extrapolation for the predictor)
polynomial. As discussed in \cite{MA92}, the fourth-order Hermite
scheme has an extra advantage that it is quite simple to
implement. The predictor polynomial for fourth-order schemes must be
at least third order for position and second order for
velocity. Fortunately, the directly calculated first time derivative
of the acceleration ({\it jerk}) is just sufficient to construct
predictors. In other words, fourth-order Hermite scheme is effectively
a single-step algorithm which does not require the memory of previous
timesteps.

Another advantage of the fourth-order Hermite scheme is that it is
time-symmetric, when used with the correct-to-convergence mode. This
feature has been used to achieve effective time-symmetry for the
integration of internal motions of binaries \citep{Funato+} or
nearly-circular orbits of planetesimals \citep{Kokubo+98}. Also, in
\citep{HMACS}, a time-symmetric individual timestep algorithm with
Hermite scheme have been implemented.

The calculation cost of the Hermite scheme per timestep is somewhat
higher than that of the Aarseth scheme, since the jerk must be
calculated as well as the acceleration. However, roughly speaking the
Hermite scheme allows the timestep larger than that for the Aarseth
scheme by almost a factor of two, while increase in the calculation
cost seems to be less than a factor of two. Thus, by switching from
the fourth-order Aarseth scheme to the fourth order-Hermite scheme,
effective gain in calculation speed is achieved while the calculation
program becomes simpler. This combined effect is the reason why the
fourth-order Hermite scheme is now widely used.

The result shown in \cite{Makino91a} implies that for blockstep
algorithms higher-order schemes might be more efficient. In this
paper, we construct higher-order generalization of the fourth-order
Hermite scheme and report their performance.

There are two different ways to construct higher-order generalization
of the Hermite scheme. The first one is to use previous timesteps, in
the same way as in the original Aarseth scheme. This method was
described in \cite{Makino91a}. The other is to use even higher
derivatives directly calculated, while still using only two points in
time. Of course, it is possible to combine these two methods.

To our knowledge, there have been no published work on the latter
approach combined with the individual timestep algorithm.  At first
sight, it looks nontrivial to combine the direct calculation of the
higher-order derivatives and individual timestep algorithm. In section
2, we show that the combination is actually possible and that it is
not much difficult compared to the original fourth-order Hermite
scheme. In section 3, we present the result of numerical experiments,
and section 4 is for discussions.

\section{Sixth- and Eighth-Order Hermite scheme}

\subsection{Basic structure of individual timestep scheme}

In the individual timestep scheme, particle $i$ has its own
time ($t_i$), timestep ($\Delta t_i$), position ($\bx_i$) and velocity
($\bv_i$) at time $t_i$, and acceleration ($\ba_i$) 
and time derivative(s) of acceleration ($\badot_i, \batwo_i, ...$) 
calculated at time $t_i$. The integration
proceeds according to  the following steps:

\begin{enumerate}

\item Select  particle $i$ with a minimum $t_i+\Delta t_i$.
Set the global time ($t$) to be this minimum, $t_i+\Delta t_i$.

\item Predict the positions and necessary time derivatives of all particles
at time $t$ using the predictor polynomials.

\item Calculate the acceleration  and its time derivative(s)
 for particle $i$ at time $t$, using the predicted positions etc.

\item Construct higher order time derivatives using the Hermite
interpolation based on the new values of acceleration and its derivatives at time
$t_i+\Delta t_i$ and those at
the previous time $t_i$. Apply the corrector to position and velocity
using these high-order
time derivatives, determine new timestep $\Delta t_i$, and update time
$t_i$. 

\item Go back to step (1).

\end{enumerate}

The above description is for the original individual timestep
algorithm, and we usually use the so-called blockstep algorithms, in
which the timesteps are quantized to powers of two so that particles
of the same stepsize share exactly the same time (McMillan 1986).  In
this way, we can calculate forces on these particles in parallel.

In the following, we present the force calculation formula, predictor,
corrector, timestep criterion and initialization procedure, in this
order. 

\subsection{Direct calculation of higher order derivatives}

The gravitational force from particle $j$ to particle $i$ and its
first three time derivatives are expressed as

\begin{eqnarray}
\bA_{ij} &=& 
	m_j \frac{\br_{ij}}{r_{ij}^3}, \label{eq:calc_acc} \\ 
\bJ_{ij} &=& 
	m_j \frac{\bv_{ij}}{r_{ij}^3} - 3\alpha\bA_{ij}, \label{eq:calc_jerk} \\ 
\bS_{ij} &=& 
	m_j \frac{\ba_{ij}}{r_{ij}^3} - 6\alpha\bJ_{ij} - 3\beta\bA_{ij}, \label{eq:calc_snap} \\
\bC_{ij} &=& 
	m_j \frac{\bj_{ij}}{r_{ij}^3} - 9\alpha\bS_{ij}
        - 9\beta\bJ_{ij} - 3\gamma\bA_{ij}. \label{eq:calc_crackle} 
\end{eqnarray}
Here, we 
call the first four time
derivatives of the acceleration {\it jerk}, {\it snap}, {\it crackle} and {\it pop}, and
$\alpha$, $\beta$ and $\gamma$ are given by
\begin{eqnarray}
\alpha &=& \frac{\br_{ij} \cdot \bv_{ij}}{r_{ij}^2}, \\
\beta  &=& \frac{|\bv_{ij}|^2 + \br_{ij} \cdot \ba_{ij}}{r_{ij}^2} + \alpha^2, \\
\gamma &=& \frac{3\bv_{ij}\cdot\ba_{ij} 
	+ \br_{ij}\cdot\bj_{ij}}{r_{ij}^2} + \alpha(3\beta - 4\alpha^2),
\end{eqnarray}
where $\br_i$, $\bv_i$, $\ba_i$, $\bj_i$ and $m_i$ are the position,
velocity, total acceleration, total jerk and mass of particle $i$, and
$\br_{ij} = \br_j - \br_i$,
$\bv_{ij} = \bv_j - \bv_i$, 
$\ba_{ij} = \ba_j - \ba_i$ and
$\bj_{ij} = \bj_j - \bj_i$ 
\citep{Aarseth2003}.

\begin{table}
\begin{center}
\caption{ Number of floating-point operations in force calculation up to
	potential, acceleration, jerk, snap and crackle.
}
\label{tab:opcount}
\begin{tabular}{l  l  r }
\hline
\multicolumn{1}{c}{Max derivative} &
\multicolumn{1}{c}{Total operations} &
\multicolumn{1}{c}{Operation count} \\
\hline
Potential    & 7 add/sub, 4 mul, 1 div, 1 sqrt & 31 \\ 
Acceleration & 10 add/sub, 8 mul, 1 div, 1 sqrt   & 38 \\
Jerk         & 21 add/sub, 19 mul, 1 div, 1 sqrt  & 60 \\
Snap         & 39 add/sub, 38 mul, 1 div, 1 sqrt  & 97 \\
Crackle      & 61 add/sub, 63 mul, 1 div, 1 sqrt  & 144 \\
\hline
\end{tabular}
\end{center}
\end{table}

Table \ref{tab:opcount} shows the number of floating point operations
needed to calculate the terms from potential to crackle.  We used the
conversion factor 20 for the operation counts for one division and one 
square root operations, following the convention introduced by \cite{WSB+}.  
Compared to the calculation
up to jerk, the increase of the operation count for higher order terms
is rather modest. Even the calculation up to crackle is only about a
factor of two more expensive than that up to jerk.  Thus, if the
eighth-order scheme allows two times larger timestep than that for
fourth order scheme for the same accuracy, the eighth-order scheme is
more efficient. Of course, the CPU time is not directly proportional
to the number of floating point operations, and therefore the actual
efficiency might be somewhat different.

\subsection{The necessary orders of predictor and corrector}
\label{sec:pred-order}

In the case of fourth-order Hermite scheme, we used two points in time
and acceleration $\ba$ and jerk $\bj$. To construct a sixth-order
scheme, we need to add one more term, snap $\bs$, to the corrector.
With two evaluations, at the beginning and at the end of the timestep,
of the three variables $\ba$, $\bj$ and $\bs$, we indeed have the 
six values needed for a sixth-order scheme.

One practical question is how to construct the predictor. In the case
of the fourth-order scheme, it is sufficient to use the terms up to
jerk, since the leading error of the predicted position then becomes
$O(\dt^4)$, which is consistent with the order of the integrator
(Aarseth 1963).  In the case of the sixth-order scheme, we need the
predictor with terms up to crackle (third derivative of the
acceleration), to be consistent. On the other hand, the corrector
requires terms only up to snap. Therefore, we directly calculate the
derivatives only up to snap, and evaluate the crackle using Hermite
interpolation.  The interpolation formula for crackle, as well as
those for fifth and sixth-order terms, are given in Appendix
\ref{app:6th}. Those for the eighth-order scheme are given in
Appendix \ref{app:8th}.

\subsection{Predictor}
\label{sec:pred}

The predictor for the sixth-order integrator is given by

\begin{eqnarray}
\br_{i,p} &=& \br_i + \bv_i\dt + \frac12\ba_i\dt^2
	+ \frac16\bj_i\dt^3 + \frac1{24}\bs_i\dt^4 
	+ \frac1{120}\bc_i\dt^5, \\ 
\bv_{i,p} &=& \bv_i + \ba_i\dt + \frac12\bj_i\dt^2
	+ \frac16\bs_i\dt^3 + \frac1{24}\bc_i\dt^4, \\ 
\ba_{i,p} &=& \ba_i + \bj_i\dt + \frac12\bs_i\dt^2 
	+ \frac16\bc_i \dt^3.
\end{eqnarray}
Note that we need to predict acceleration, since it is used to
calculate snap (see equation \ref{eq:calc_snap}).
For eighth-order scheme, we need to include two additional terms for
each predictor, and we also need to predict jerk. Since the predictor
is simply a Taylor expansion, we do not give the specific forms for
the eighth-order scheme here.

\subsection{Corrector}

The sixth-order corrector is given by

\begin{eqnarray}
\bv_{i,c} &=& \bv_{i,0} 
	+ \frac{\dt}{2}(\ba_{i,1} + \ba_{i,0})
	- \frac{\dt^2}{10}(\bj_{i,1} - \bj_{i,0})
	+ \frac{\dt^3}{120}(\bs_{i,1} + \bs_{i,0}),\\
\br_{i,c} &=& \br_{i,0} 
	+ \frac{\dt}{2}(\bv_{i,c} + \bv_{i,0})
	- \frac{\dt^2}{10}(\ba_{i,1} - \ba_{i,0})
	+ \frac{\dt^3}{120}(\bj_{i,1} + \bj_{i,0}).
\end{eqnarray}

See Appendix \ref{app:6th} for the details of derivation. Note that we
gave the simplest form for the corrector of the position, which
uses jerks but not snaps. It is possible to construct the corrector
which use the snaps, but that would
not change the order of the time integration. For special problems
such as integration of near-Kepler orbit with constant timestep,
appropriate treatment of this highest-order term improves the behavior
of the integrator \citep{Kokubo+04}.

The eighth-order corrector is
given by
\begin{eqnarray}
\bv_{i,c} = \bv_{i,0} 
	&+& \frac{\dt}{2}(\ba_{i,1} + \ba_{i,0})
	- \frac{3\dt^2}{28}(\bj_{i,1} - \bj_{i,0}) \nonumber \\
	&+& \frac{\dt^3}{84}(\bs_{i,1} + \bs_{i,0})
	- \frac{\dt^4}{1680}(\bc_{i,1} - \bc_{i,0}), \\
\br_{i,c} = \br_{i,0} 
	&+& \frac{\dt}{2}(\bv_{i,c} + \bv_{i,0})
	- \frac{3\dt^2}{28}(\ba_{i,1} - \ba_{i,0}) \nonumber \\
	&+& \frac{\dt^3}{84}(\bj_{i,1} + \bj_{i,0})
	- \frac{\dt^4}{1680}(\bs_{i,1} - \bs_{i,0}).
\end{eqnarray}

See \ref{app:8th} for the details of the derivation.

\subsection{Timestep criterion}
\def\acck#1{\ba^{(#1)}}
\def\absak#1{|\acck{#1}|}

For a high-order integration scheme with adaptive timestep to work
properly, it is essential to use an appropriate timestep criterion. In
this paper we consider two different timestep criteria. The first one
is the generalization of the ``Aarseth'' criterion
\begin{equation}
\dt = \sqrt{\eta\frac{ 
	|\ba||\ba^{(2)}| + |\ba^{(1)}|^2}{
	|\ba^{(1)}||\ba^{(3)}| + |\ba^{(2)}|^2}},
\label{eq:aarseth}
\end{equation}
where  $\acck{k}$ is the $k$th derivative of acceleration  
and $\eta$ is a parameter which controls the accuracy. Usually, a
value around 0.02 is used for $\eta$.
This criterion is known to work well with fourth-order schemes, but it
is also known that it works well only with fourth-order schemes and
does not give good results for higher-order schemes
\citep{Makino91a}.
\cite{Aarseth2003} notes that this criterion should be generalized
to include the highest derivative available in higher-order integrators.

We tried to generalize the above criterion to higher-orders as
\begin{equation}
\dt = \eta \left(\frac{A^{(1)}}{A^{(p-2)}}
\right)^{1/(p-3)}
\label{eq:generalaarseth}
\end{equation}
where
\begin{equation}
A^{(k)} = \sqrt{\absak{k-1}\absak{k+1} + \absak{k}^2}.
\label{eq:generalaarsethdef}
\end{equation}
Here, $p$ is the order of the integrator.  We moved the accuracy
parameter $\eta$ out of the fractional power, so that the timestep is
directly proportional to $\eta$.  The numerator is the same as that
for the Aarseth criterion for the fourth-order scheme, and for the
denominators we used the terms of highest orders available. The
fractional power is chosen to give correct dimension of time.  This
criterion should behave reasonably well, since it does reflect the
high-order terms.

We also tested a criterion which is based on  the error of the predictor
\begin{equation}
\Delta t_{new} = \Delta t_{old} \left(
		\frac{\epsilon|\ba|}{|\ba - \ba_p|}
		\right)^{1/p},
\label{eq:dadt}
\end{equation}
where $\ba_p$ is the predicted acceleration and $\ba$ is calculated one.
In order to use this criterion  the acceleration needs to be predicted
to the highest order
 (see section \ref{sec:pred-order}).

\subsection{Initialization}

One practical advantage of the fourth-order Hermite scheme is that it
is effectively a single-step algorithm and therefore does not need any
special initialization procedure, except that the initial timestep
must be chosen differently.  Unfortunately, this single-step nature is
lost when we go to higher orders, since higher derivatives for the
predictor need to be constructed using a Hermite interpolation.

The simplest implementation of the initialization procedure is just to
use lower-order predictors for the first timestep and use
appropriately small timestep. In our current implementation, for the
startup of the sixth-order integrator we use the terms up to crackle
directly calculated and the Aarseth criterion (\ref{eq:aarseth}), as
was done in the Aarseth code. Thus, the order of the predictor is
consistent. For the eighth-order integrator, we omit the calculation
of further derivatives and  start with (\ref{eq:aarseth}) with a
small accuracy parameter.

\section{Numerical test}

In this section, we report the behavior of the sixth- and eighth-order
Hermite schemes, for time integration of a 1024-body Plummer model. We
used the standard units where the total mass of the system and
gravitational constants are both unity and the total energy of the
system is $-1/4$. For all calculations, we used a softened
gravitational potential with $\varepsilon=4/N = 1/256$.  We used the
block timestep algorithm 
timesteps are restricted to be powers of two, and we set an upper
limit of timestep as 1/16.
All calculations were performed in IEEE 754 double precision.

\subsection{Result for short-time integration}

\begin{figure}
\begin{center}
\includegraphics[height=\hsize, angle=270, clip]{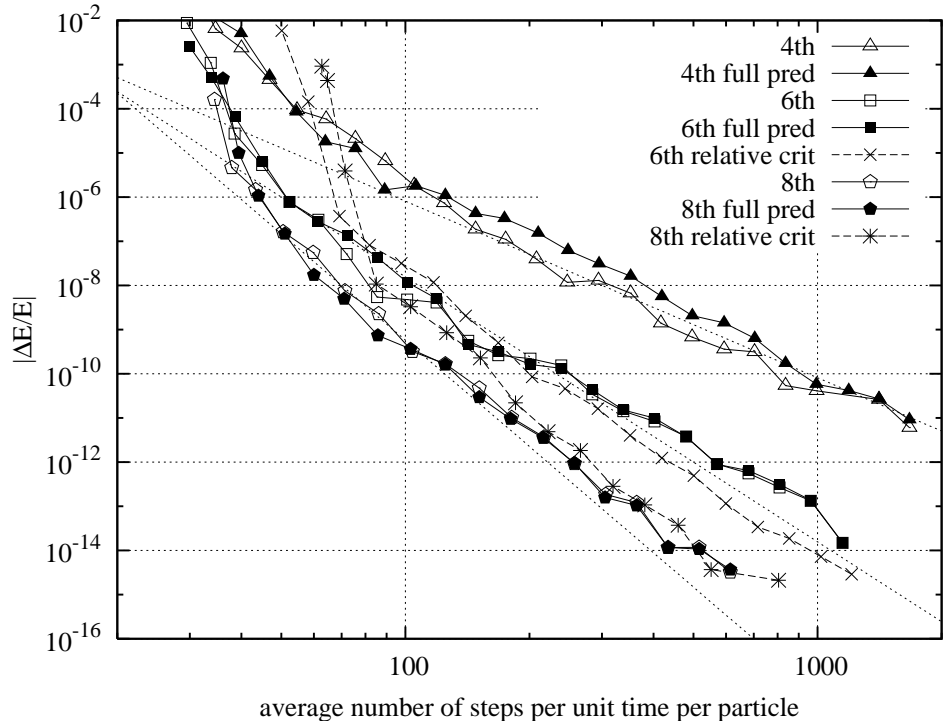}
\caption{
Maximum relative deviation of the total energy during the time integration for
10 time units, as a function of average number of timesteps per
particle per unit time.  Triangles, squares and pentagons represent
the results of 4th-, 6th- and 8th-order schemes. Open, filled and star-shape
symbols indicate the lowest order predictor with generalized Aarseth
criterion (\ref{eq:generalaarseth}), full high-order predictors with
generalized Aarseth criterion, and full high-order predictor with
timestep criterion based on the predictor error (\ref{eq:dadt}).
The three dotted lines indicate the expected scaling relations for 
4th-, 6th- and 8th-order algorithms (top to bottom).
}
\label{fig:error}
\end{center}
\end{figure}

Figure \ref{fig:error} shows the relation between the maximum relative
energy error during the integration for 10 time unit and the average
number of timesteps per particle per unit time.  To suppress possible
effects of the start-up procedure, we first integrate the system for
$1/8$ unit time and measured the maximum relative deviation of the
energy during the next 10 time units.

We can clearly see that the error of sixth- and eighth-order schemes
are proportional to $\dt^6$ and $\dt^8$, as expected. For the relative
accuracy of $10^{-8}$, the sixth-order scheme allows the average
timestep which is almost a factor of three larger than that necessary
for the fourth-order scheme. For the relative accuracy of $10^{-11}$,
the eights-order scheme allows the average timestep which is about a
factor of seven larger than that necessary for the fourth-order
scheme. Even for the relatively low accuracy of $10^{-6}$, the sixth-order
scheme allows about a factor of two larger timestep than the
fourth-order scheme does.

Among the three schemes (different predictor orders and different
timestep criteria), the difference in the achieved accuracy is not
very large, but there are some trends. For example, when we compare
the results with low-order predictors and that with high-order
predictors, low-order predictors give systematically better results,
at least for the fourth order integrators.  One possible reason is
that the high-order predictor is effectively less time-symmetric
compared to low-order predictor, simply because it uses the
information from the previous timestep.  When we compare two timestep
criteria, formula (\ref{eq:dadt}) seems to be worse, at least for
large stepsizes. Thus, among these three implementations, the
combination of low order predictor and generalized Aarseth criterion
seems to be the most safe. It also requires the least amount of
floating point operations per timestep.

\subsection{Long-term integration}
\label{sec:long}

We integrated the system until the core collapse occurs. Since we
used a softened potential, a compact core of the size comparable to
the softening length is formed after the core collapse. We stopped the
calculation at that point.  The accuracy parameter was chosen so that
the actual CPU time per unit time integration is initially
similar. The actual value of the accuracy parameter for the criterion
(\ref{eq:generalaarseth}) is 0.1, 0.4, and 0.75 for 4, 6 and 8th order
schemes.

Notice that the system has a chaotic nature, and even if we start from
slightly different two models, the positions of each particle would
be completely different after a long-term integration.

\begin{figure}
\begin{center}
\includegraphics[height=1.0\hsize, angle=270, clip]{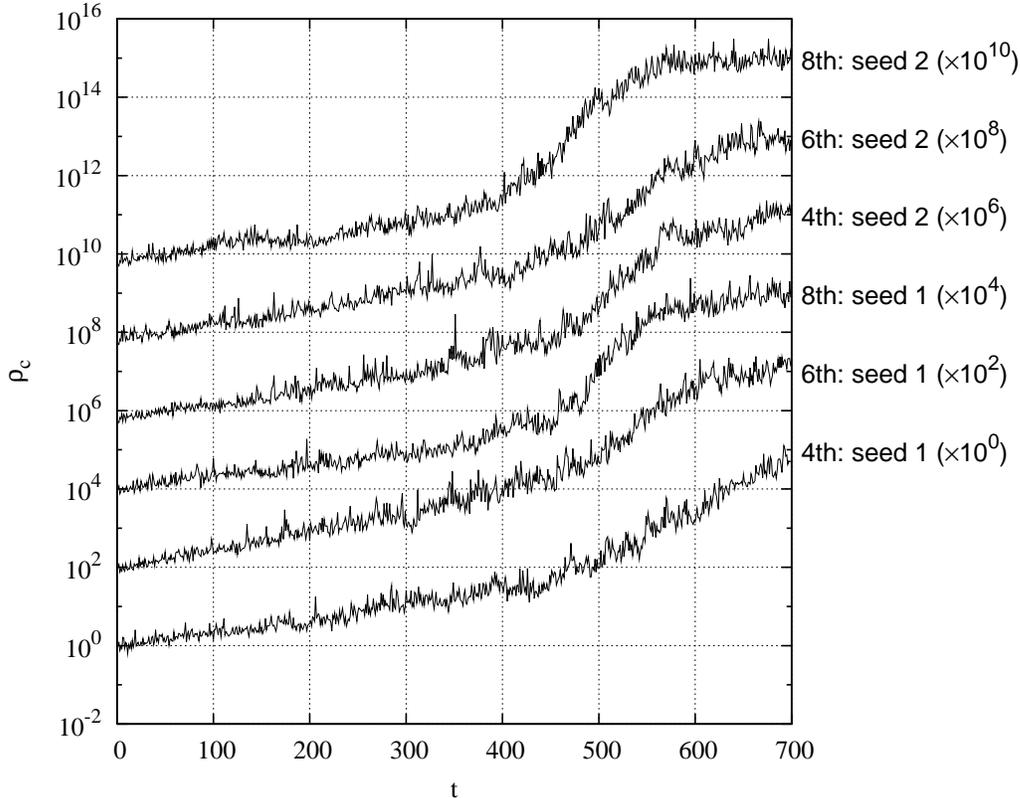}
\caption{ Time evolution of the central density $\rho_c$ for long-term
  integrations, for three different integration orders and two
  different values of the random seed for the initial model. The
  curves are shifted by factors of 100.
}
\label{fig:dence}
\end{center}
\end{figure}
\begin{figure}
\begin{center}
\includegraphics[height=\hsize, angle=270, clip]{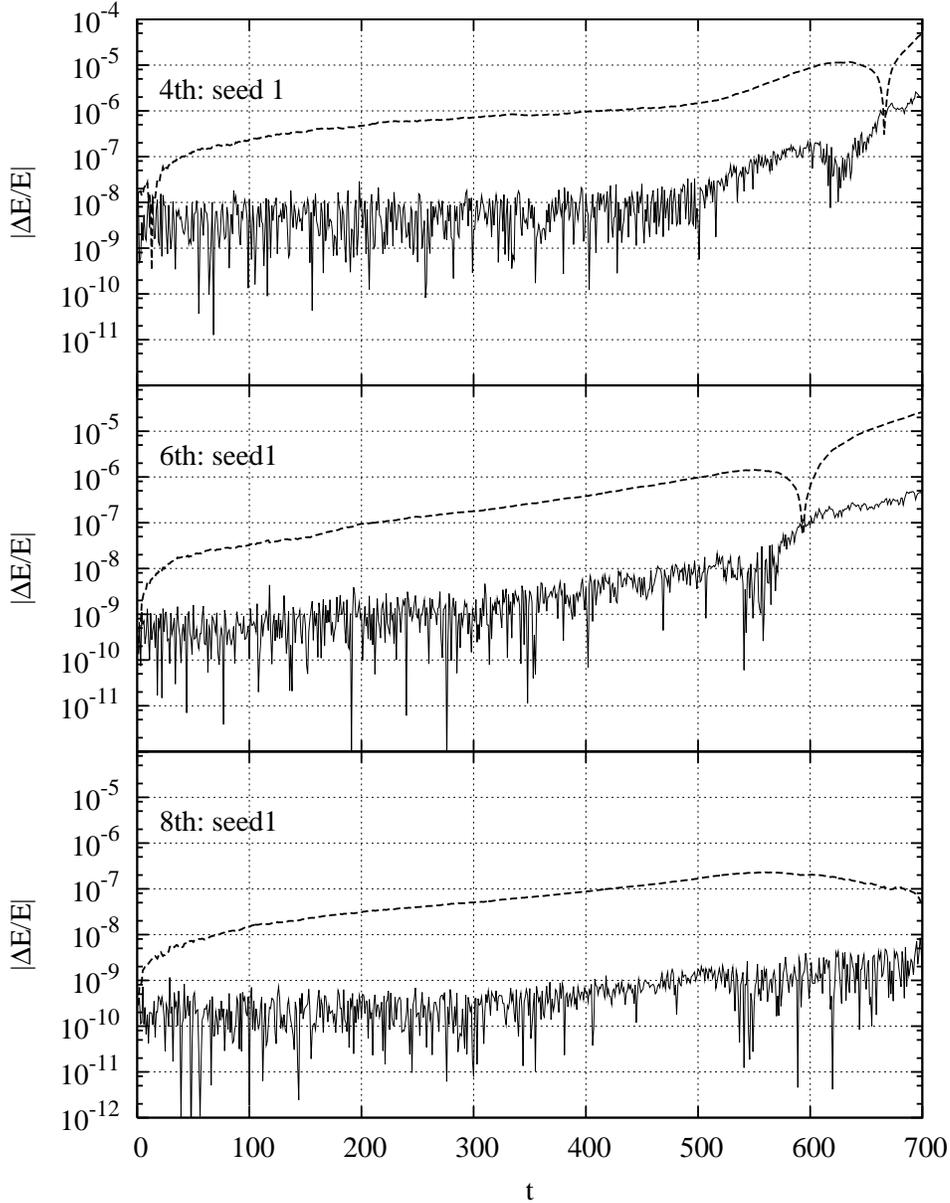}
\caption{
Total energy error ($|E(t)-E(0)|/|E(0)|$, dashed curve) and
energy change in one unit time ($|(E(t)-E(t-1)|/|E(0)|$, solid curve)
in 4th, 6th and 8th order integrators.  
}
\label{fig:de}
\end{center}
\end{figure}
\begin{figure}
\begin{center}
\includegraphics[height=\hsize, angle=270, clip]{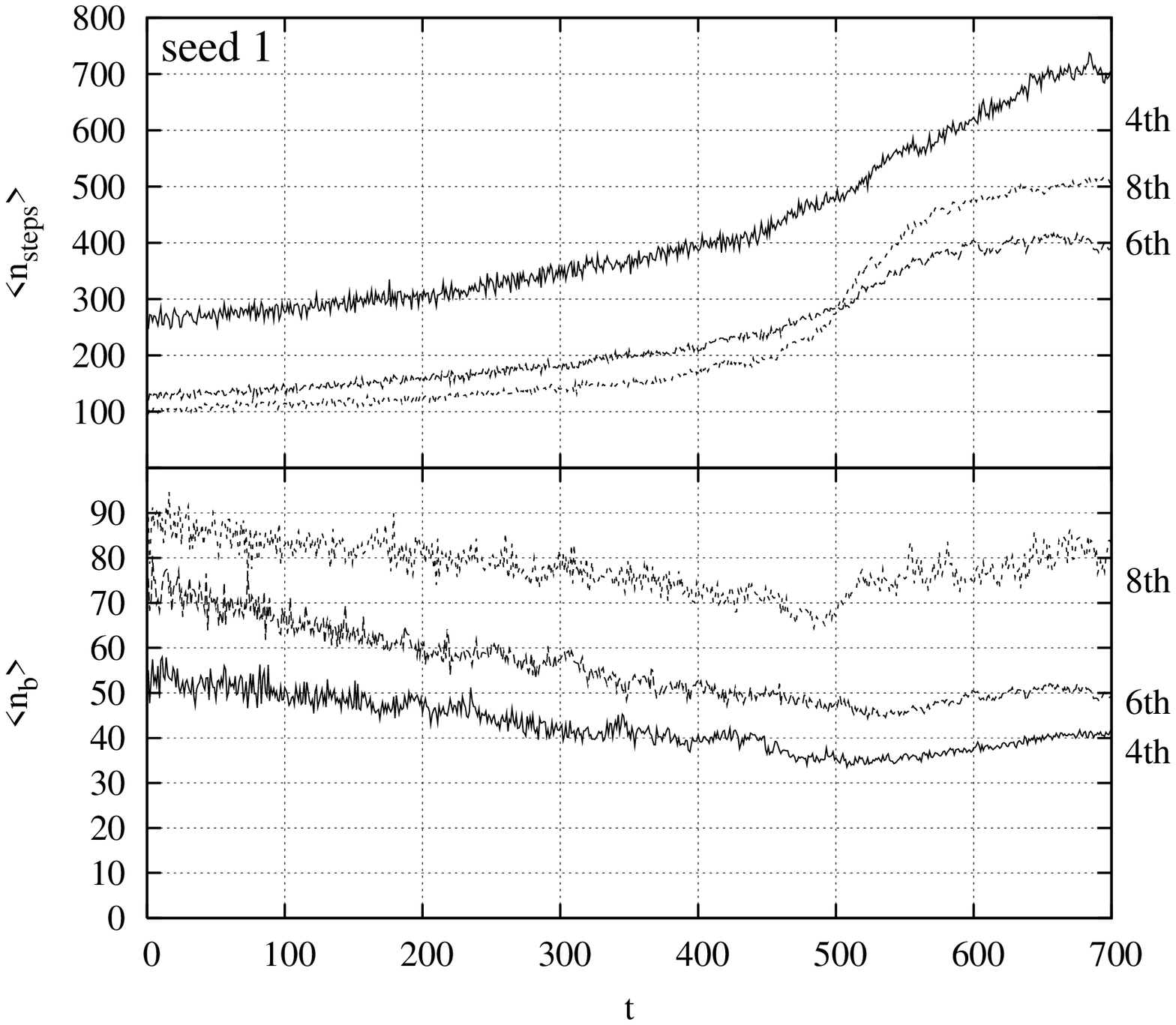}
\caption{
	The average number of steps per particle per unit time $<n_{steps}>$ (top)
	and the average number of particles advanced in one block-step $<n_b>$ (bottom). 
}
\label{fig:nstep}
\end{center}
\end{figure}

Figure \ref{fig:dence} shows the evolution of the central density $\rho_c$. 
The overall behavior is similar for all runs. 
Figure \ref{fig:de} shows the cumulative relative energy error and 
relative energy error per unit time.
As expected, higher-order schemes achieve higher accuracy, though the
number of timestep per unit time is smaller.
With all schemes, the error per unit time becomes larger as the system
evolves. However, this  increase is smaller for higher-order
schemes. If we compare the error per unit time at the beginning and
the end of the calculation, with fourth-order scheme the error at the
end is bigger by nearly three orders of magnitude. In the case of the
eighth-order scheme, the increase is only around a factor of 10. In
the case of the sixth-order scheme, the increase is in between those
for fourth- and eighth-order schemes.

This result seems to suggest that the higher-order schemes are
actually more robust than the fourth-order scheme. However, it is not
clear from where  such a difference comes from. Naively, the
increase of the integration error is understood as coming from
particles in the core, which need to be integrated for a large number
of orbits since the orbital timescale is short. Since the structure of
the system is not much different, orbital timescales of particles in
the core should not depend on the integration schemes used, and there
is no reason for the errors to behave differently. 

Figure \ref{fig:nstep} shows the average number of timesteps per
particles per unit time $<n_{steps}>$ (top panel) and the average
number of particles integrated in one blockstep $<n_b>$ (bottom
panel). We can see that the increase in the number of timesteps is the
largest for the eighth-order scheme, and this increase in the number
of timesteps seems to be the reason for the small error in the later
time shown in figure \ref{fig:de}. Another notable behavior of the
higher-order schemes is that the average number of particles
integrated in one blockstep is larger in higher-order schemes. This
means that the timestep criteria for Hermite schemes with different
orders respond differently to the change of the structure of the
system.  In the following, we examine the reason of this behavior of
the timesteps.

\begin{figure}
\begin{center}
\includegraphics[height=0.9\hsize, angle=270, clip]{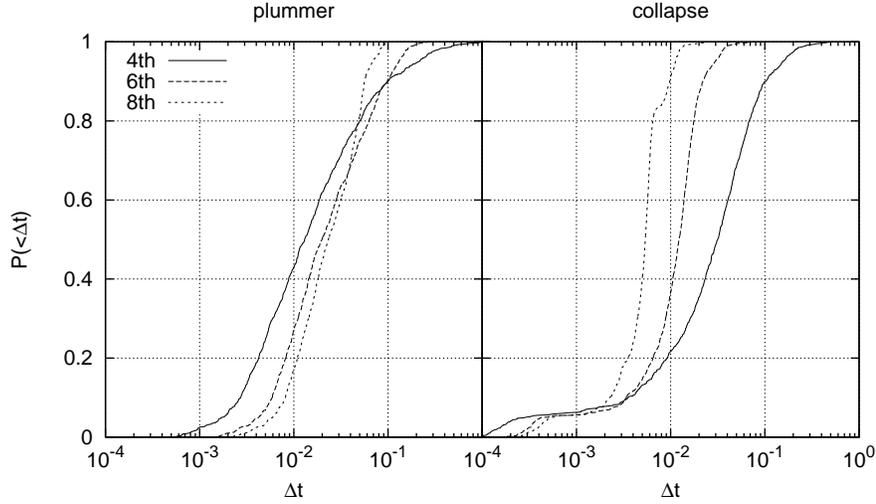}
\end{center}
\caption{
	Cumulative distribution of timestep for the initial Plummer model (left)
	and that after the core collapse (right). Solid, dashed and
dotted curves shows the timesteps calculated with  fourth-, sixth-,
and eighth-order timestep criteria. 
}
\label{fig:dtfrac}
\end{figure}

\begin{figure}
\begin{center}
\begin{tabular}{c  c}
\includegraphics[height=0.5\hsize, angle=270, clip]{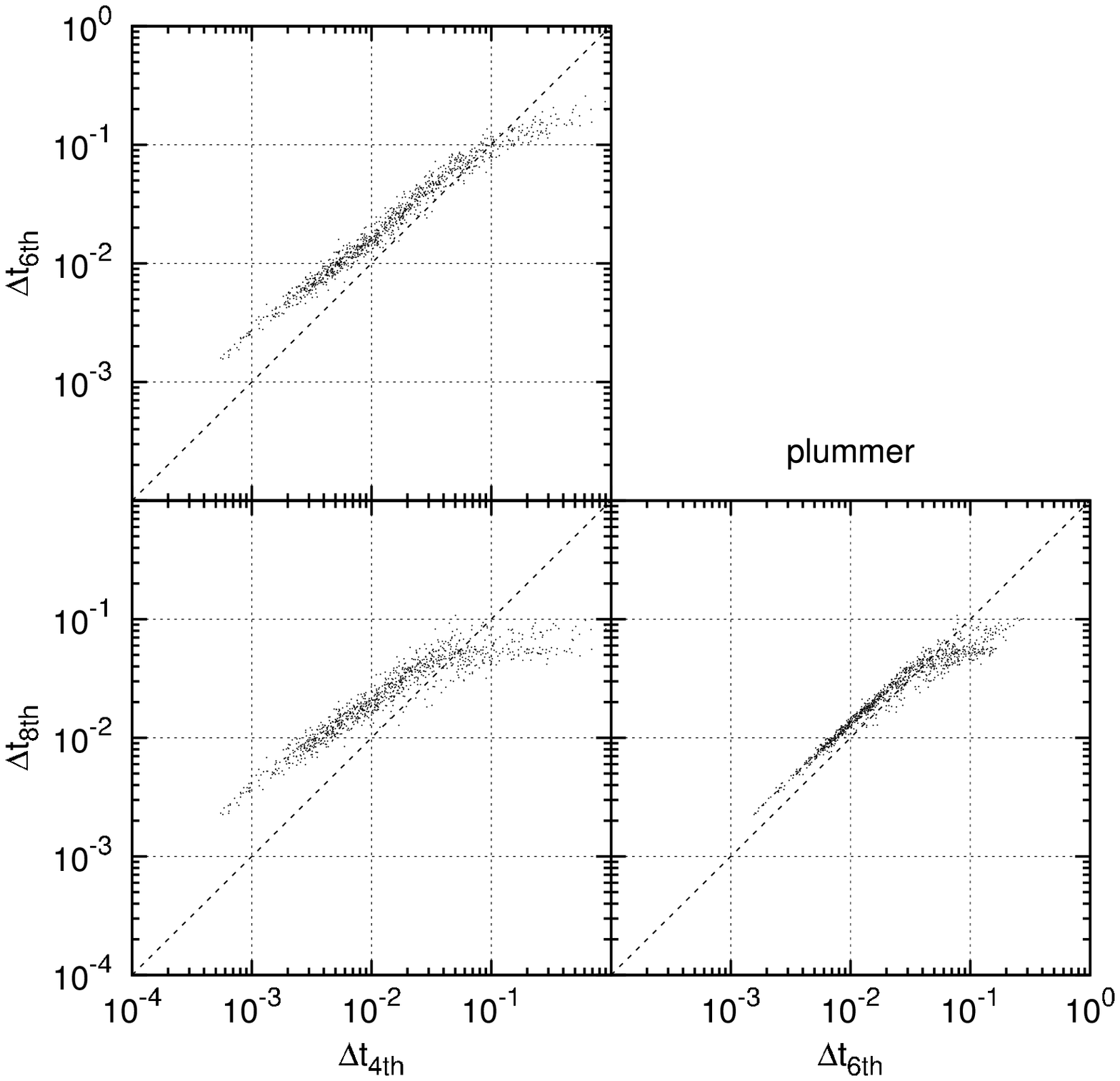} &
\includegraphics[height=0.5\hsize, angle=270, clip]{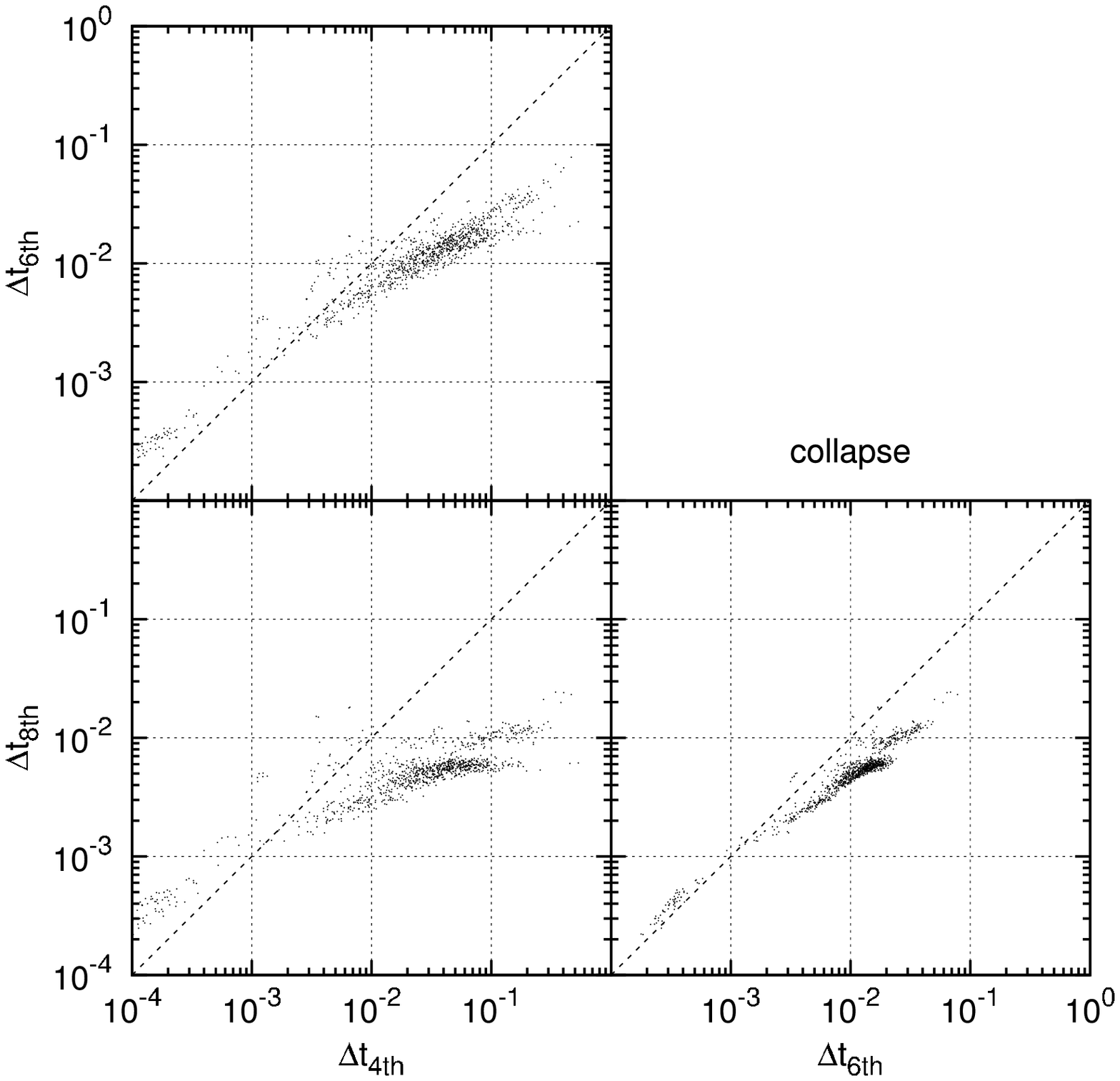} \\
\end{tabular}
\end{center}
\caption{
	Relations of timesteps calculated using the timestep criteria
for three integrations schemes with  orders, 
	for the initial Plummer model (left) and after
	the core collapse (right). Top-left, bottom-left and
bottom-right panels shows the relations between schemes with orders
(4-6), (4-8) and (6-8), respectively.
}
\label{fig:dtrel}
\end{figure}

In order to examine the behavior of the timestep criterion for
different orders, it is necessary to calculate the timesteps with
different orders for identical distribution of particles. We can of
course do this for initial condition, but exact comparison is
impossible for the later times. In order to make the comparison, we
used the system integrated with the eighth-order scheme, and
calculated the timesteps with fourth- and sixth-order criteria using
the derivatives obtained by the eighth-order Hermite interpolation.
To determine the stepsize, we used the accuracy parameter $\eta$ same
as that used in the actual time integration.

Figure \ref{fig:dtfrac} shows the distribution of timesteps for
initial Plummer model and after the core collapse ($T=700$).
We can see that the distribution of timesteps depends strongly to the
order of the integrator. Even for the initial condition, the range of timesteps
of particles for the eighth-order scheme is much narrower than that for
the fourth-order scheme. This tendency is much more pronounced for the
system after the core collapse. Thus, for particles with smallest
timesteps, timestep criteria with  different orders give similar
values. However, particles with large timesteps have very different
stepsizes, depending on the order of the timestep criterion.

Figure \ref{fig:dtrel} shows actual values of timesteps calculated
using timestep criteria of different orders as two-dimensional
scattergrams.  We can see the correlation is quite tight, but at the
same time highly nonlinear. While particles with small stepsize (those
in the central region of the system) have similar stepsizes when
calculated with different orders, particles with large stepsizes
(those in the outer region of the system) have very different stepsize
depending on the order. In particular, it seems high-order schemes
have maximum stepsizes which depend on the structure of the system.
The initial Plummer model allows timesteps up to
around 0.1, while for the collapsed system timestep cannot
significantly exceed 0.01 in the case of the eighth-order scheme.
In other words, the timestep of particles far away from the central
region of the system somehow shrinks by a factor of 10 in the case of
the eighth-order criterion, while no such shrinking is visible for
the fourth-order criterion.

\subsection{Toy model}

In order to understand this behavior, let us consider a simplified
model of the collapsed system. Assume that the distribution of
particles outside the core is isothermal with $\rho \propto r^{-2}$, and
the core size is $r_c$. Total mass of the system within radius $r=1$
is 1, and orbital timescale of particles at radius $r$ is $r$. The
mass inside radius $r$ is $r$ for $r_c<r<1$. The number of particles
in $r<1$ is $N$ and mass of particles $m$ is $1/N$. In these units, The core mass
$m_c$ is around $r_c$, within a factor of two or so.

First, we consider the contribution of the nearest neighbor, for a
particle at $r=1$. The
distance to the nearest neighbor is roughly $r_n = 1/N^{1/3}$. The
timescale in which this distance changes is also $(r_n/v)=1/N^{1/3}$,
where $v$ is the typical relative velocity and is order unity.  Therefore,
the acceleration and its $k$-th derivative have the strength of around
\begin{equation}
|a_n^{(k)}| \sim m r_n^{-2} (r_n/v)^{-k} \sim N^{(k-1)/3}.
\end{equation}
We can see that the timestep
criteria of the form (\ref{eq:generalaarseth}) would give similar
stepsize of $1/N^{1/3}$ for different orders, if the higher order terms
are dominated by the contributions from near neighbors.
To put it in words: if you use a local criterion to determine an 
integration step time scale, you will always get the time needed 
to reach your nearest neighbor.

Now consider the contribution from one particle at small distance 
$r<<1$ from the center, to a particle at distance one from the center. 
The strength of the acceleration is of the order $m=1/N$.
The orbital timescale is the larger of  $r$ and $r_c$. We call this
value, the larger of $r$ and $r_c$, as  $r_*$. 
The timescale of the change of the acceleration $\Delta t$ is the orbital
timescale $r_*$, and the fractional change in the acceleration
is also order $r_*$ : $\Delta a \sim m r_*$. 
Thus the time derivatives of the acceleration  have the strength of
\begin{equation}
|a_r^{(k)}| \sim \Delta a / (\Delta t)^k 
            \sim m r_* \cdot r_*^{-k} 
			\sim \frac{1}{N r_*^{k-1}} \quad (k >0).
\label{eq:corejerk}
\end{equation}

If $ r_* < 1/N^{1/3}$, 
the contribution to the high order
derivative of a particle at distance one from  the center of the
system is larger for a particle with distance $r \le r_*$
from the center than for the nearest neighbor, for sufficiently large
values of $k$.
For our numerical experiment, the size of the core at the core
collapse is around 0.003, which is much smaller than $1/N^{1/3}$. Thus,
for high enough orders like $k=7$, it seems natural that the
contributions from the particles in the core dominate. For $k=3$, the
nearest neighbor might still be dominant.

From the viewpoint of the accuracy of the time integration, it seems
obvious that the behavior of higher order criteria is better. With
low-order criteria, we effectively ignore the high-frequency variation
of the acceleration when integrating the orbits of particles far away
from the core. As a result, there must be the sampling error
or aliasing error for the forces from particles in the core, which
should show up as the integration error. The energy error of one
particle, due to one particle in the core, in one timestep would be of
the order of $\Delta E \sim mv\Delta a \Delta t$, since the distance one
particle moves in one timestep is $ mv \Delta t$ and the force error
applied during the time the particle moves is $\Delta a$. Thus, we
have $\Delta E \sim \Delta t r_c N^{-2}$.

The number of error terms is proportional to the number of 
particles in the system $N$, number of particles in the core $r_cN$, 
and the average number of timesteps per unit time $\Delta t^{-1}$.
Thus, for one time unit, the total error would be of the order of
\begin{equation}
\Delta E_{\rm sampling} 
	\sim \sqrt{N \cdot r_cN \cdot \Delta t^{-1}} \Delta t r_c N^{-2}
	=    N^{-1} r_c^{2/3} \Delta t^{1/2}.
\end{equation}
Here, we assumed that the errors are random and the total error is
proportional to the square-root of the number of error terms.
Either assumption might not be correct. For example, error terms of
the forces on different particles from one particle in the central
region are likely to be correlated. Therefore the actual error might
be larger. 

In the above, we assumed that the error comes only from the sampling
of the acceleration. However, with Hermite schemes, what we actually
integrate is the high order derivatives directly calculated. Thus, the
actual error is probably much bigger than the estimate above. In the
case of the fourth-order Hermite scheme which uses the first time
derivative, the derivative is of the order $1/N$ from equation
(\ref{eq:corejerk}). Therefore, the energy error of one particles 
is more like $\Delta E \sim mv|a_r^{(1)}| \Delta t^2 \sim \Delta t^2 N^{-2}$ 
per timestep, and total error is of order of
\begin{equation}
\Delta E_{\rm sampling, jerk} \sim N^{-1} r_c^{1/2} \Delta t^{3/2}.
\end{equation}

Note that the only way, for the current individual timestep scheme,
to make this error reasonably small is to
shrink the timestep of {\it all} particles in the system so that the
orbits of core particles are resolved. In that sense, the behavior of
the eighth-order scheme is numerically correct, and that of
fourth-order scheme is not. This is probably the reason why the error
increases for the fourth-order scheme.

\section{Discussion}
\subsection{Computational aspects}

In this paper, we present sixth- and eighth-order
Hermite schemes to be used with individual timestep algorithm, and
compared their performance with that of the fourth-order scheme. We
found that these higher-order schemes do offer practical advantages.

Here we speculate on the merit 
of higher-order schemes,
when used with special-purpose computers or parallel implementation of
individual timestep algorithm on massively-parallel
computers.

One advantage of the direct calculation of higher-order derivatives in
hardware is that calculation of higher-order terms can be implement
with short word lengths. In the case of GRAPE-4 \citep{Grape4} or
GRAPE-6 \citep{Grape6}, the calculation of acceleration is done with
24-bit mantissa and jerk with 20-bit mantissa. Similarly, it would be
okay to calculate snap in 16 bits and crackle in 12 bits. Thus, though
the number of operation becomes large as we calculate higher-order
terms, the silicon area needed does not increase much. Even in the
case of general-purpose programmable computers (including GPUs),
higher-order schemes have extra advantage, since the calculation of
high-order terms can be done in lower precision, for example single
precision. Many of general-purpose CPUs have extra instructions for
fast single-precision operations, and GPUs are much faster in single
precision than in double precision. 

Another advantage is that the number of particles integrated in one
blockstep is larger for higher-order schemes. Thus, we can enjoy more
parallelism with higher-order schemes, resulting in higher execution
efficiency  for both general-purpose parallel computers and GRAPEs.

Thus, the higher-order schemes are  better not just in the pure
operation count but also in hardware efficiency and parallel
efficiency.

\subsection{Physical/mathematical aspects}

In section \ref{sec:long}, we have seen that, for high-order schemes,
timesteps of a particles in outer region of the system become short to
resolve the high-frequency variation of the acceleration due to
particles in the central region. In the case of low-order schemes,
such shrinking did not occur. However, as a result the energy error
became very large. Thus, it seems that we need to make the timestep of
all particles small, as the system evolves and the core size becomes
smaller. This means that the analysis of the calculation cost based on
the assumption that the nearest neighbor determines the timestep
\citep{MH88} is not correct.

Our conclusion implies that the individual timestep algorithm is not
so effective as shown in \citet{MH88}, because the maximum timestep of
particles is larger than that of the shortest timestep by the factor
which only weakly depends on the number of particles in the system. In
principle, however, we can reduce the calculation cost by assigning
individual timesteps not to particles but to interactions. Even when
there is a small core with short orbital period, the force between two
particles both far away from the core changes smoothly. Strictly
speaking, it is not really smooth, since the position and velocity
contain the high-frequency terms. However, these high frequency terms
are much smaller than that of accelerations simply because position
and velocity are obtained by integrating the acceleration. Thus, if we
use relatively low-order schemes or the original Adams-type scheme
which does not use the time derivatives, we might be able to integrate
the interaction between two particles far away from the core with the
timestep much larger than that of core particles.  We will investigate
the possibility of such a scheme in future papers.

\medskip
We thank Sverre Aarseth for his lectures and discussions at Mitaka
on November, 2007. 
K.~N.~ is financially supported by Research Fellowship of the
Japan Society for the Promotion of Science (JSPS) for Young Scientists.
This work was supported in part by the Special Coordination Fund for 
Promoting Science and Technology (GRAPE-DR project), from the Ministry 
of Education, Culture, Sports, Science and Technology (MEXT) of  Japan.

\appendix

\section{Notes for implementers}
\subsection{Interpolation polynomial for the sixth-order integrator}
\label{app:6th}
We consider constructing an interpolation polynomial from
acceleration, jerk and snap at time $t_0$, $(a_0, j_0, s_0)$
and those at $t_1$, $(a_1, j_1, s_1)$.
We define summations and differences of them as
\begin{eqnarray}
A^+ &\equiv& {a_1 + a_0}, \nonumber\\
A^- &\equiv& {a_1 - a_0}, \nonumber\\
J^+ &\equiv& h(j_1 + j_0), \nonumber\\
J^- &\equiv& h(j_1 - j_0), \nonumber\\
S^+ &\equiv& h^2(s_1 + s_0), \nonumber\\
S^- &\equiv& h^2(s_1 - s_0),
\end{eqnarray}
where $h=(t_1-t_0)/2$ and $a^{(k)}$ is the $k$th derivative of $a$.
Coefficients of the interpolation polynomial at the midpoint $t=(t_0+t_1)/2$ are

\begin{eqnarray}
                a_{1/2}       &=& \frac1{16} (8A^+ - 5J^- + S^+), \nonumber \\
              h j_{1/2}       &=& \frac1{16} (15A^- - 7J^+ + S^-), \nonumber \\
\frac{h^2}{2}   s_{1/2}       &=& \frac18 (3J^- - S^+), \nonumber \\
\frac{h^3}{6}   a^{(3)}_{1/2} &=& \frac18 (-5A^- + 5J^+ - S^-), \nonumber \\
\frac{h^4}{24}  a^{(4)}_{1/2} &=& \frac1{16} (-J^- + S^+), \nonumber \\
\frac{h^5}{120} a^{(5)}_{1/2} &=& \frac1{16} (3A^- - 3J^+ + S^-).
\end{eqnarray}
By shifting them to $t_1$, we have derivatives for the next predictor or timestep criterion.
\begin{eqnarray}
a^{(3)}_1 &=& a^{(3)}_{1/2} + h a^{(4)}_{1/2} + \frac{h^2}{2} a^{(5)}_{1/2}, \nonumber \\
a^{(4)}_1 &=& a^{(4)}_{1/2} + h a^{(5)}_{1/2}, \nonumber \\ 
a^{(5)}_1 &=& a^{(5)}_{1/2}.
\end{eqnarray}
By integrating the even-order terms, we obtain the sixth-order corrector
\begin{eqnarray}
v_1 - v_0 &=& h \left(
	A^+ - \frac25 J^- + \frac1{15} S^+
\right).
\end{eqnarray}

\subsection{Interpolation polynomial for the eighth-order integrator}
\label{app:8th}
The interpolation polynomial for the eighth-order integrator
is constructed from $(a_0, j_0, s_0, c_0)$ and $(a_1, j_1, s_1, c_1)$.
Here $c$ is crackle.
Even-order coefficients are
\begin{eqnarray}
\begin{array}(r)
			 a^{}_{1/2} \\
	\dfrac{h^2}{2}   s^{}_{1/2} \\
	\dfrac{h^4}{24}  a^{(4)}_{1/2} \\
	\dfrac{h^6}{720} a^{(6)}_{1/2}
\end{array}
	&=& \frac{1}{32}
\begin{array}({rrrr})
	16 & -11 &  3 & -1/3 \\
	 0 &  15 & -7 &    1 \\
	 0 &  -5 &  5 &   -1 \\
	 0 &   1 & -1 &  1/3
\end{array}
\begin{array}(c)
	A^+ \\
	J^- \\
	S^+ \\
	C^-
\end{array},
\end{eqnarray}
and odd-order ones are
\begin{eqnarray}
\begin{array}(r)
			  h\ j^{}_{1/2} \\
	\dfrac{h^3}{6}    c^{}_{1/2} \\
	\dfrac{h^5}{120}  a^{(5)}_{1/2} \\
	\dfrac{h^7}{5040} a^{(7)}_{1/2}
\end{array}
	&=& \frac{1}{32}
\begin{array}({rrrr})
	35 & -19 &   4 & -1/3 \\
   -35 &  35 & -10 &    1 \\
	21 & -21 &   8 &   -1 \\
	-5 &   5 & - 2 &  1/3
\end{array}
\begin{array}(c)
	A^- \\
	J^+ \\
	S^- \\
	C^+ 
\end{array},
\end{eqnarray}
with
\begin{eqnarray}
C^+ &\equiv& h^3(c_1 + c_0), \nonumber\\
C^- &\equiv& h^3(c_1 - c_0).
\end{eqnarray}
The eighth-order corrector is
\begin{eqnarray}
v_1 - v_0 &=& h \left(
	A^+ - \frac{3}{7} J^- + \frac{2}{21} S^+ - \frac{1}{105} C^-
\right).
\end{eqnarray}


\end{document}